\documentclass[prb, reprint, superscriptaddress, showpacs]{revtex4-1}
\usepackage{amsmath, amsfonts, amssymb, bm}
\usepackage{color, graphicx}
\usepackage[colorlinks,citecolor=blue, urlcolor=blue,linkcolor=blue]{hyperref}

\begin{document}

\title{Phase diagram of the Hubbard model on the anisotropic triangular lattice}

\author{Manuel Laubach}
\author{Ronny Thomale}
\affiliation{\mbox{Institut f\"ur Theoretische Physik und Astrophysik,
  Universit\"at W\"urzburg, 97074 W\"urzburg, Germany} }
\author{Christian Platt}
\affiliation{\mbox{Department of Physics, Stanford University, Stanford, California 94305, USA} }
\affiliation{\mbox{Institut f\"ur Theoretische Physik und Astrophysik,
  Universit\"at W\"urzburg, 97074 W\"urzburg, Germany} }
\author{Werner Hanke}
\author{Gang Li}
\affiliation{\mbox{Institut f\"ur Theoretische Physik und Astrophysik,
  Universit\"at W\"urzburg, 97074 W\"urzburg, Germany} }

\begin{abstract}
We investigate the Hubbard model on the anisotropic triangular lattice
as a suggested effective description of
the Mott phase in various triangular organic compounds.
Employing the variational cluster approximation
and the ladder dual-fermion approach as complementary
methods to adequately treat the zero-temperature and the finite-temperature domains, we obtain a consistent picture of the phase
diagram as a function of anisotropy and interaction strength.    
The metal-insulator transition
substantially depends on the anisotropy, and so does the nature of
magnetism and the emergence of a nonmagnetic insulating phase.
We further find that geometric anisotropy significantly influences the
thermodynamics of the system. For increased frustration induced by
anisotropy, the entropy of the system 
increases with interaction strength, opening the
possibility of adiabatically cooling a frustrated system by an
enhancement of electronic correlations.
\end{abstract}

\date{\today}

\maketitle

\section{Introduction}
Single crystals of organic charge-transfer salts have recently received substantial interest, 
and fascinating phenomena such as several emergent many-body phases have been 
observed.
$\kappa\mbox{-(BEDT-TTF)}_{2}\mbox{Cu[N(CN)}_{2}\mbox{]Cl}$ and
$\kappa\mbox{-(BEDT-TTF)}_{2}\mbox{Cu}_{2}\mbox{(CN)}_{3}$ are two
prototypical examples of such organics: As far as similarities are concerned,  
the Fermi liquid, Mott insulator behavior, and the crossover between the
two phases are 
similarly detected in both materials~\cite{PhysRevLett.85.5420,
  PhysRevLett.95.177001}.
The metal-insulator transition is found to be associated with
unusual critical exponents which, face value, do not fall into any
established universality class~\cite{Nature436.2005}.  
The main difference between these compounds, however, which sparks
even more substantial interest, shows up in their magnetic behavior.
$\kappa\mbox{-(BEDT-TTF)}_{2}\mbox{Cu[N(CN)}_{2}\mbox{]Cl}$ 
displays long-range magnetic order at low temperatures, which is in 
sharp contrast to the strongly
magnetically frustrated behavior of
$\kappa\mbox{-(BEDT-TTF)}_{2}\mbox{Cu}_{2}\mbox{(CN)}_{3}$.
The bulk spin susceptibility~\cite{PhysRevLett.91.107001} of
$\kappa\mbox{-(BEDT-TTF)}_{2}\mbox{Cu}_{2}\mbox{(CN)}_{3}$ reflects
this strong frustration, as it displays no
indication of long-range antiferromagnetic order at
temperatures significantly lower than the magnetic exchange scale, as inferred from the
Heisenberg coupling estimated 
from high-temperature series
expansions~\cite{PhysRevB.71.134422}.  
Together, these findings highlight the similarity of the two compounds in the
charge sector,  and also the clear difference regarding the spin degrees of freedom,
which also manifests itself in
the effective-field-theory description of the
problem~\cite{PhysRevB.76.235124,PhysRevLett.102.176401,PhysRevB.79.064405, 
PhysRevLett.95.036403}. 
Separating charge from spin paves the way for the investigation of effective
spinon theories in 
the Mott insulating phase, with or without gapless spinon modes
yielding a potentially unstable spinon Fermi surface~\cite{maissam}.

Such effective field theories, however, are not stringently specified
and can take on different forms. For example, the low-$T$ thermal
conductivity is found to be contributed by the spin-1/2 spinons in a
theory of a U($1$) gauge field coupled to a spinon Fermi
surface~\cite{PhysRevB.76.235124}, but is associated with visons in a 
${\cal Z}_{2}$ theory~\cite{PhysRevLett.102.176401}.   
A puzzling situation similar to that in theory likewise exists on the
experimental side. There, the interpretation of the given evidence is
far from settled, as specific-heat~\cite{Nature.4.459} and thermal-conductivity~\cite{Nature.5.44} experiments may
yield different conclusions on the nature of the fermionic excitations in 
$\kappa\mbox{-(BEDT-TTF)}_{2}\mbox{Cu}_{2}\mbox{(CN)}_{3}$. 
On top of all these complications, even if we assume a spin liquid state in
the latter compound, it is still debated whether this state would be
fully gaped or not~\cite{doi:10.1146/annurev-conmatphys-062910-140521,0034-4885-74-5-056501}. 

What are the microscopic parameters whose variations impose such a
diversity of exotic many-body phenomena in the organic compounds?  
While the interaction strength is probably rather comparable in
all these compounds, a clear
difference that catches the eye lies in the anisotropy
strength inherently determined by the underlying crystal structure and
chemical components.  The anisotropy strength can be obtained from
{\it ab initio} calculations~\cite{PhysRevLett.103.067004,
  JPSJ.78.083710, 2012arXiv1208.3954N}.
Assuming the Hubbard model on a triangular
lattice is the correct model to describe the interplay of geometric 
frustration with electronic correlations, it is a
natural further step of complexity to consider the lattice anisotropy.

In this paper, we study the Hubbard model on the triangular lattice with varying 
anisotropy strengths, devising methods to treat both the
zero- and finite-temperature regime. The Hamiltoinan is given by
\begin{equation}
  \hat{H}= -\sum_{<ij>,\sigma}t_{ij} 
  c^{\dagger}_{i\sigma}c^{}_{j\sigma}-\mu
  \sum_{i\sigma}c_{i\sigma}^{\dagger}c_{i\sigma} +
  U\sum_i n_{i \uparrow}n_{i \downarrow},
  \label{eq:ham_anisotrop}
\end{equation}
with $\left<ij \right>$  denoting nearest neighbor bonds, 
$t_{ij}=t'$ for the horizontal hopping, and $t_{ij}=t$ for the diagonal
hopping as shown in Fig.~\ref{fig:anisotropic-lattice}. 
Varying the anisotropy $t'/t$ from the limiting values 0 to 1, we 
effectively change the lattice geometry from square to triangular
type. Unless stated otherwise in the paper, we choose the phrasing of
small and large anisotropy according to the value of $t'/t$. The most
interesting regime for the organic compounds is located around the
isotropic triangular limit $t'/t=1$, where only small
variations can yield crucially different scenarios. It is 
instructive to see how the phase diagram evolves from the square
limit towards the triangular limit, which is why we analyze the
complete domain $0 \le t'/t \le 1.1 $. For $t'/t>1.1$, the system
quickly evolves towards an effectively 
one-dimensional scenario of weakly coupled chains, quickly
rejecting magnetic 
order~\cite{ronnyjohannes}.  This is neither a relevant regime for the organic
compounds we focus on, nor particularly suited for the methods employed in this
work, which is why the regime $t'/t>1.1$ will not be further addressed.

This paper is organized as follows. Section~\ref{Sec:Method} briefly introduces
the methodology of the variational cluster approximation (VCA) and the ladder dual-fermion approach (LDFA). For the VCA, we elaborate on
several recent methodological refinements to enhance its quantitative
accuracy such as treating the hopping as a variational parameter and avoiding artificial broadening to obtain the exact poles of the single-particle Green's function. The zero- and
finite-temperature phase diagrams are obtained by 
VCA and LDFA, respectively, providing a complementary and consistent 
perspective of the Hubbard model on the anisotropic triangular lattice. Our results
are discussed in Sec.~\ref{Sec:Results}. As a function
of the anisotropy parameter from the square to the triangular limit, we find a magnetic phase transition from
N\'eel-antiferromagnetic (AFM) to 120$^{\circ}$-AFM  order along with a growing metallic
regime for weaker Hubbard $U$. Furthermore, already close to the triangular limit, the
onset of a nonmagnetic insulating regime is found, which is the
candidate domain for possible spin-liquid states, where the charge
degree is frozen without spin ordering. The LDFA additionally offers
the possibility of addressing questions of thermodynamics in the
Hubbard model. In particular, we find an indication of adiabatic cooling
caused by the change of frustration as a function of interaction
strength, which might be observed in highly tunable scenarios such as
triangular optical lattices loaded with ultra-cold fermionic isotopes. 
In Sec.~\ref{Sec:Conclusions}, we conclude that our analysis sets the
initial stage for further investigations of 
the many-body phases in the Hubbard model on the anisotropic
triangular lattice, for which we can identify the promising non-magnetic insulating
regime. Whatever unconventional phases may be found in
this regime, and whichever effective theories best describe them, the lattice
anisotropy is likely to be a crucial microscopic parameter.

\section{Methodology}\label{Sec:Method}
In this section, we briefly review the methods we employ in this
paper, namely, the VCA and the LDFA. 
These two methods will be subsequently applied
for zero and finite temperature, respectively, focusing on the quantum
phase diagram and certain thermodynamic properties. 

\subsection{$T=0$: Variational cluster approach}
The VCA~\cite{Potthoff2003b} is based on the self-energy-functional theory
(SFT)~\cite{Potthoff2003a,Potthoff2005}, which provides an efficient
numerical technique for studying strongly correlated systems, especially in the presence of different competing orders.   
VCA simplifies the lattice problem, as defined in
Eq.~(\ref{eq:ham_anisotrop}), to an exactly solvable problem
defined in a reference system consisting of decoupled finite-size
clusters. 
The thermodynamic limit is recovered by reintroducing the intercluster hopping
to the decoupled cluster  via a nonperturbative
variational scheme based on SFT. 
The VCA has been successfully applied to many interesting problems,
including the high-$T_{c} $ cuprates~\cite{dahnken-2004-70,
  PhysRevLett.94.156404, 0295-5075-72-1-117, Hanke2010, Brehm2010} and
topological insulators~\cite{PhysRevLett.107.010401,Laubach2014a,
  PhysRevB.83.041104}. 

In particular, the VCA has already been
employed to analyze the Hubbard model on the anisotropic triangular
lattice by Sahebsara and S\'en\'echal~\onlinecite{PhysRevLett.97.257004}. It is a method that is particularly suitable for such a study, as the anisotropy induces several phase transitions in the geometrically frustrated system which can be conveniently described within VCA. In Ref.~\onlinecite{PhysRevLett.97.257004}, however, the
specifications chosen within the VCA, such as the choice of the
finite-size cluster, were inadequate to correctly resolve a
significant range of the phase diagram.
In the following, we will present our VCA in an
independent and self-contained fashion. The most important refinements
we employ for the VCA to obtain an accurate phase diagram, as
well as a detailed comparison to
Ref.~\onlinecite{PhysRevLett.97.257004}, are explicated in the Appendix.

In the SFT, the grand potential of a system is defined by
$H=H_0(\mathbf{t})+H_1(\mathbf{U})$ and written as
a functional of the self-energy $\Sigma$: 
\begin{align}
  \Omega[\Sigma]&= F\left[ \Sigma \right]+\text{Tr}\ln\left(
  G^{-1}_0-\Sigma \right)^{-1} \, , 
  \label{eq:grand-potential-functional}
\end{align}
where $F\left[ \Sigma \right]$ is the Legendre transform of the
Luttinger-Ward functional and $G_0=(\omega+\mu-\mathbf{t})^{-1}$ is
the non-interacting Green's function.
It can be shown that the functional $\Omega[\Sigma]$ is
stationary at the physical self-energy, i.e., $\delta\Omega\left[
  \Sigma_{\text{phys}} \right]=0$~\cite{Potthoff2003a}. 
As the Luttinger-Ward functional is universal, it has the
same interaction dependence for systems with any set of single-particle
operators $\mathbf{t'}$ 
as long as the interaction $\mathbf{U}$ remains unchanged. 
Note that the functional $\Omega\left[ \Sigma \right]$ itself is not
approximated by any means; 
we restrict, however, the ``parameter'' space of possible
self-energies to the self-energies of the reference system. 
Thus, the stationary points are obtained from the self-energy
$\Sigma'=\Sigma\left[ \mathbf{t'} \right]$ of a system defined by 
$H^{\prime}=H_0(\mathbf{t'})+H_1(\mathbf{U})$, which we call the reference
system. 
After defining $V=\mathbf{t}-\mathbf{t}'$, we are able to conveniently
define the VCA Green's function, 
\begin{equation}
G_{\rm VCA}^{-1} = G'^{-1} - V\ .
\end{equation}
The VCA grand potential is
\begin{align}
  \Omega[\Sigma']&= \Omega'+\text{Tr}\ln\left( G^{-1}_0-\Sigma'
  \right)^{-1} - \text{Tr}\ln(G') \, ,
  \label{eq:grand-potential}
\end{align}
with $\Omega'$, $\Sigma'$, and $G'$ denoting the grand potential, the
self-energy, and the single-particle Green's function of the reference
system, respectively. 
The reference system is chosen such that it can be solved exactly; that is, $\Omega'$, $\Sigma'$, and $G'$ should be readily obtained numerically. 
We choose the reference system as a set of decoupled
clusters and solve them with open boundary conditions via exact
diagonalization.   
In this sense, the short-range correlations within the reference
system are fully taken into account in the VCA. 
The correlation beyond the reference system size will be treated in a
mean-field fashion via the variational scheme. 

The choice of the reference system, {\it i.e.}, the cluster
shape and size, has to respect the fact that tuning the
anisotropy from 0 to 1 effectively modifies the system geometry from
square to triangular. 
Thus, a reference system is needed which incorporates both the N\'eel and spiral
orders in the VCA in a commensurate fashion in order not to bias the
variational procedure (see also App.~\ref{SC-app}).  
\begin{figure}
  \begin{center}
    \includegraphics[width=0.8\linewidth]{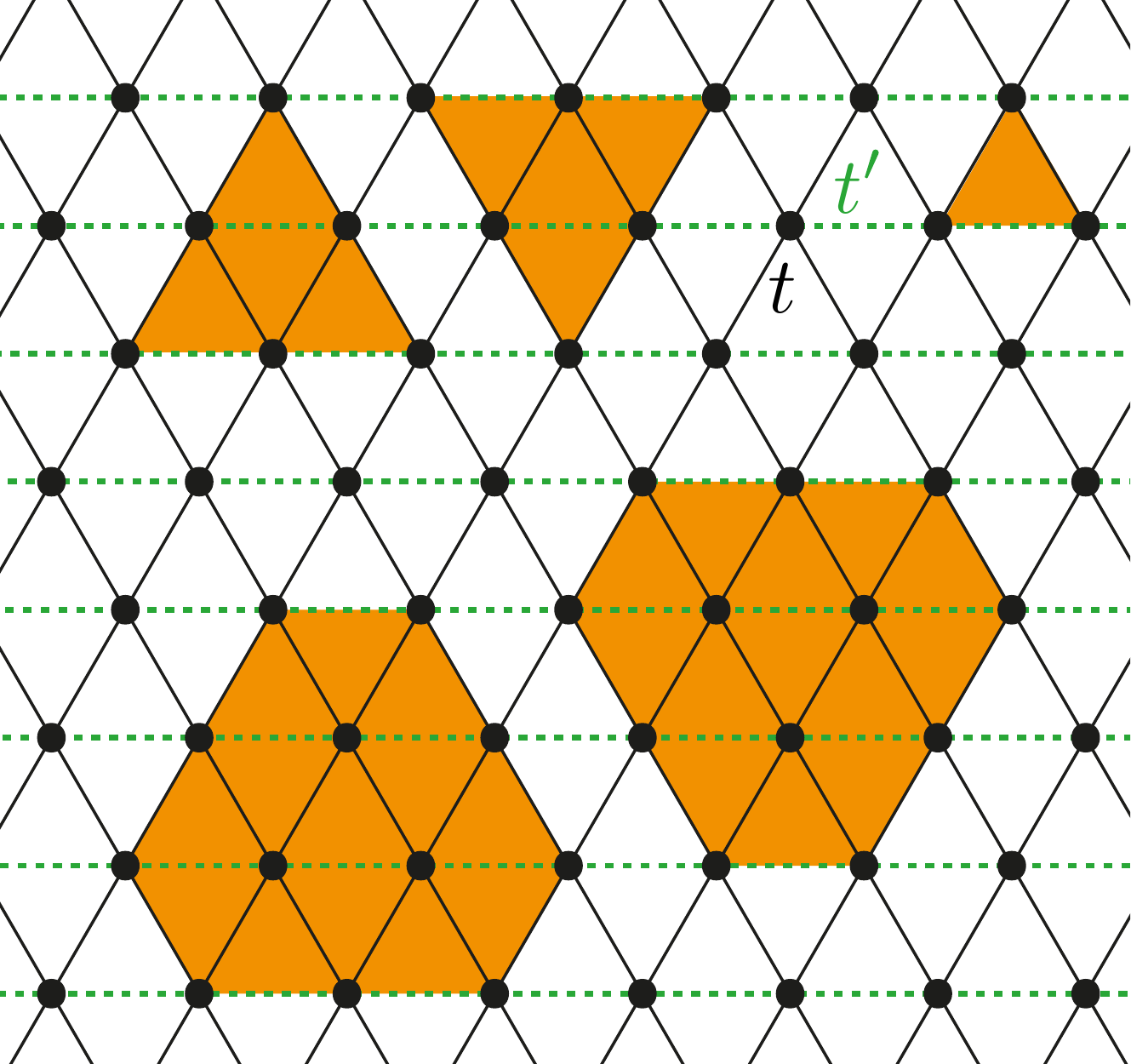}
  \end{center}
  \caption{Anisotropic triangular lattice with diagonal hoppings
    $t$(black) and vertical hopping $t'$(green). The reference clusters
    are shown for $L_c=3$, $L_c=6$, and $L_c=12$, where the larger clusters are 
    mirrored to recover the lattice geometry.
    Hoppings between the cluster and its mirror cluster are indicated by the dotted links.}
  \label{fig:anisotropic-lattice}
\end{figure} 
In this paper, two reference systems are explored with cluster
sizes of $L_{c}=6$ and $L_{c}=12$.
They are mirrored in our calculation to recover the lattice translation
symmetry with a supercluster~\cite{Sahebsara2008}.
The Green's function of this supercluster $G'$ consists of the cluster
and mirrored cluster given as, 
\begin{eqnarray}  \label{eq:cluster-green-mit-mirror}
  G'^{-1}=\left( 
  \begin{matrix} 
    G_1^{\prime-1} & 0\\ 
    0 & G_2^{\prime-1} 
  \end{matrix} \right) + \left( 
  \begin{matrix}
    0 & t_{21} \\ 
    t_{12}& 0 
  \end{matrix} \right)\, ,
\end{eqnarray}
where $G_1'$ is the reference-cluster Green's function and
$G_2'$ is the Green's function of the mirrored cluster, which is a
simple transformation of $G_1'$ (in the simplest case, it is just a
copy of $G_{1}^{\prime}$). 
The reference and the mirrored clusters are connected through the
single-particle hopping $t_{12}$, as indicated by the dotted links in
Fig.~\ref{fig:anisotropic-lattice}. 

\subsection{Finite-$T$: Ladder dual-fermion approach}
\begin{figure}[bbb]
\centering
\includegraphics[width=\linewidth]{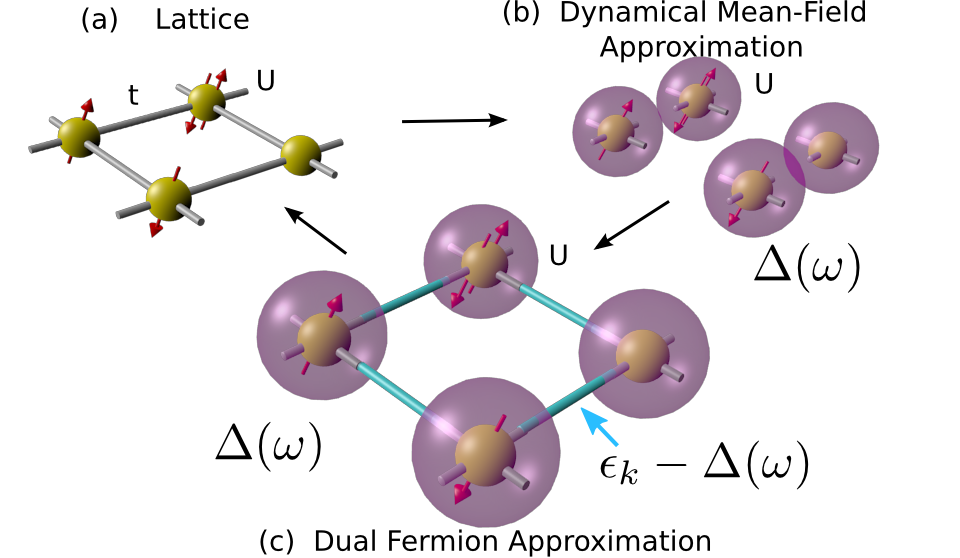}
\caption{As a non-local extension of the DMFT, the dual-fermion method
perturbatively expands the difference of the single-particle hopping and the DMFT
hybridization function, {\it i.e.}, $\epsilon_{k}-\Delta(i\omega_{n})$,
which generates systematic nonlocal corrections to 
the DMFT solution. Ideally, with all the expansion terms taken into
account, the lattice problem defined in (a) can be exactly solved by the dual-fermion method.} 
\label{DF_Graphic}
\end{figure}

For the finite-temperature study, we employ the
dual-fermion approach~\cite{PhysRevB.79.045133,
  PhysRevLett.102.206401}, considering only the two-particle vertex and
ladder-type diagrams for the self-energy.  
The dual-fermion approach decouples a correlated lattice defined in
Eq.~(\ref{eq:ham_anisotrop}) into a group of impurities which couple to
each other through an effective interaction mediated by auxiliary
dual fermions.   
The local problem can be well described within the dynamical
mean-field theory (DMFT)~\cite{RevModPhys.68.13}. 
The perturbation expansion over the effective interaction of the
dual-fermion variables can systematically generate non-local
corrections to the DMFT.  
The basic idea of the dual-fermion approach is schematically shown in
Fig.~\ref{DF_Graphic}, where the lattice problem defined in
Fig.~\ref{DF_Graphic}(a) is decoupled into an impurity problem as in
DMFT [Fig.~\ref{DF_Graphic}(b)]. The difference between the
lattice and the decoupled impurity problem is treated perturbatively in
the dual-fermion approach [Fig.~\ref{DF_Graphic}(c)].

Let Fig.~\ref{DF_Graphic}(a) denote the Hubbard model on a square
lattice, where the yellow spheroids represent sites on the lattice. The
bond connecting two yellow spheroids represents the hopping between
these two sites. When two electrons with different spins stay on the
same site, local Hubbard $U$ acts upon them.  
In the DMFT approximation [Fig.~\ref{DF_Graphic}(b)], the bonds between
different sites are effectively removed, in the sense that each site
becomes an impurity  coupled to a dynamical bath $\Delta(\omega)$. In
Fig.~\ref{DF_Graphic}(b), $\Delta(\omega)$ is shown as a purple sphere
around each site.  $t_{ij}-\Delta(\omega)$ associated with the blue
bonds in Fig.~\ref{DF_Graphic}(c) represents the difference between the lattice model in
Fig.~\ref{DF_Graphic}(a) and its DMFT approximation in
Fig.~\ref{DF_Graphic}(b). The dual-fermion method then performs a perturbative
expansion in terms of this difference, and as such restores momentum
dependence starting from the DMFT limit. 

Let us formulate the above idea by starting with the lattice action corresponding to the Hamiltonian in 
Eq.~(\ref{eq:ham_anisotrop}): 
\begin{equation}\label{Eq:full-action}
S[c, c^{*}] = \sum_{i}S_{loc}^{i}[c, c^{*}] +
\sum_{n,k,\sigma}[\epsilon_{k} -
  \Delta(i\omega_{n})]c_{k\sigma}^{*}c_{k\sigma}. 
\end{equation}
$S_{loc}^{i}[c, c^{*}] =
-\sum_{n,\sigma}c_{k\sigma}^{\dagger}(i\omega_{n})[i\omega_{n}+\mu -
  \Delta(i\omega_{n})]c_{k\sigma}(i\omega_{n})
+ U\int n_{i\uparrow}(\tau)n_{i\downarrow}(\tau)d\tau$ is the action
of an impurity coupled to a continuum bath. The dynamics of the bath
is described by the hybridization function $\Delta(i\omega_{n})$. 
The second sum in Eq.~(\ref{Eq:full-action}) is the term that is
treated perturbatively.

If the local action $S_{loc}$ is already a good description of the
original system, the second term on the right-hand-side of Eq. (\ref{Eq:full-action}) 
will effectively be a small parameter. 
An expansion in this small term, {\it i.e.}
$\epsilon_{k}-\Delta(i\omega_{n})$, generates further corrections to 
$S_{\text{loc}}^{i}$ and can be calculated order by order.  
A convenient way for such an expansion is to rewrite the second term on
the right-hand-side of Eq. (\ref{Eq:full-action}) with a dual variable $f,
f^{*}$ through Gaussian integration.
After integrating out the $c$ variables,  the original lattice problem described in
Eq. (\ref{Eq:full-action}) now can be equally written as an
action that depends on only the $f$ variables: 
\begin{widetext}
\begin{equation}\label{Eq:new-action}
S[f, f^{*}] = -\sum_{n,k}\ln
[\Delta(i\omega_{n})-\epsilon_{k}]-\sum_{i}\ln{\cal
  Z}_{loc}^{i} + \sum_{i}V_{i}[f^{*},
f]+\sum_{n,k,\sigma}\left\{[\Delta(i\omega_{n})-\epsilon_{k}]^{-1} + g(i\omega_{n})\right\}f_{\omega k\sigma}^{*}f_{\omega k\sigma} \, .  
\end{equation}
\end{widetext}
The interaction between the $f$ variables, {\it i.e.}, $V_{i}$ in
Eq.~(\ref{Eq:new-action}), is the reducible multiparticle vertex of the $c$ variables, which can also be
obtained by solving $S_{\text{loc}}^{i}$ with an appropriate impurity solver,
such as the continuous-time quantum Monte Carlo
method~\cite{PhysRevB.72.035122}.    

Equations~(\ref{Eq:full-action}) and ~(\ref{Eq:new-action}) are two
equivalent ways to describe the same problem, as no approximation
is introduced in the transformation. Thus, the lattice Green's
function $G_{k}(i\omega_{n})$ can be equally constructed from this new
action:
\begin{equation}\label{Eq:exact-1}
G_{k}(i\omega_{n}) =
[\Delta(i\omega_{n})-\epsilon_{k}]^{-2}g_{k}^{d}(i\omega_{n}) +
[\Delta(i\omega_{n})-\epsilon_{k}]^{-1}  \, ,
\end{equation}
where  $g^{d}_k(i\omega_{n})$ is given as  
\begin{equation}
g_{k}^{d}(i\omega_{n}) =
[g^{-1}(i\omega_{n})+\Delta(i\omega_{n})-\epsilon_{k} -
  \Sigma^{d}_{k}(i\omega_{n})]^{-1}   \, .
\end{equation}
It becomes transparent that the problem of solving an interacting
many-body problem defined in Eq.~(\ref{eq:ham_anisotrop}) is
equivalent to solving an Anderson impurity problem, 
{\it i.e.}, self-consistently determining $g(i\omega_{n})$ and
$\Delta(i\omega_{n})$, and additionally calculating
$\Sigma_{k}^{d}(i\omega_{n})$ from the perturbation expansion of 
$V_{i}[f^{*}, f]$ in Eq.~(\ref{Eq:new-action}). 
In the following calculation, we will impose the approximation to consider ladder-type
diagrams of $\Sigma_{k}^{d}(i\omega_{n})$ up to infinite
order which contain only the two-particle
vertex~\cite{PhysRevLett.102.206401} in the particle-hole channel. 
With such a simplification employed, the dual-fermion approach is now
denoted as LDFA in the following. 
The LDFA approximation~\cite{PhysRevLett.102.206401}, and even the
stronger approximation of considering only selective self-energy
diagrams from the two-particle vertex~\cite{PhysRevB.78.195105} have
proven fairly accurate in studying strongly correlated electron systems.
 
Equation (\ref{Eq:exact-1}) sets up an exact relation for the
single-particle Green's function of the original $c$ variables and
the dual $f$ variables.  
Similar exact relations can also be found for higher-order correlators.
For example, for the spin susceptibility, which is employed to
identify different magnetic phases in this work, we have
\begin{eqnarray}\label{Eq:exact-2}
\chi_{Q}(k,k^{\prime}) = \chi_{Q}^{0}(k,k^{\prime}) +
h_{k}h_{k+Q}\tilde{\chi}_{Q}^{d}(k,
k^{\prime})h_{k^{\prime}}h_{k^{\prime}+Q} \, . \nonumber \\
\end{eqnarray}
Here, $h_{k}=[\Delta(i\omega_{n}) - \epsilon_{k}]^{-1}$ and
$\tilde{\chi}_{Q}^{d}(k,
k^{\prime})=\chi^{d}_{Q}(k,k^{\prime})-\chi^{d,0}_{Q}(k,k^{\prime})$
stands for the reducible vertex of the dual variables. 
The high momentum resolution of the spin susceptibility calculated
from the dual-fermion approach is very hard to achieve in other
approaches. 
In turn, this resolution is vital to studying the spin structure at different
anisotropy strengths, as the magnetic order changes from N\'eel-type
[with magnetic wave vector $Q=(\pi, \pi)$] to a $120^{\circ}$-type [with
$Q=(2\pi/3, 2\pi/3)$].

\section{Results}\label{Sec:Results}

\subsection{VCA}

\begin{figure}[bbb]
\centering
\includegraphics[width=\linewidth]{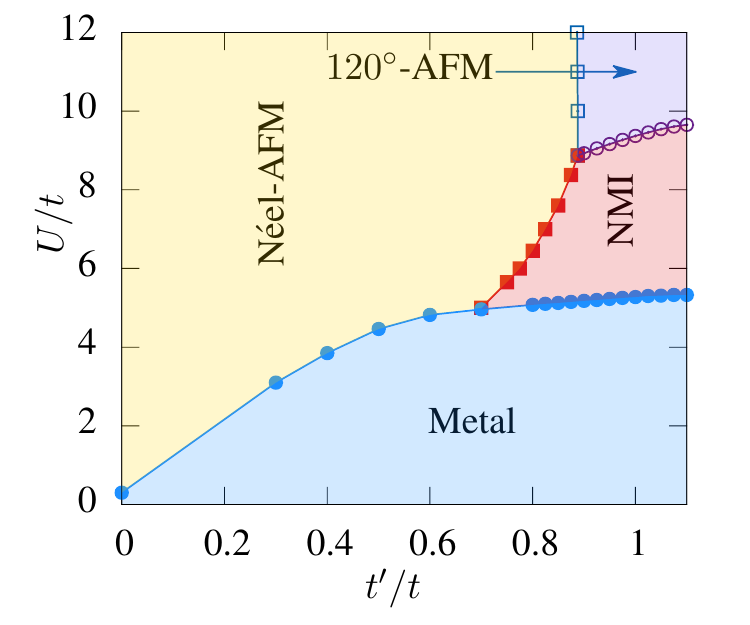}
\caption{Zero-temperature VCA phase diagram as a function of
  anisotropy $t'/t$ and interaction strength $U/t$. Changing $t^{\prime}/t$ from 0 to 1 interpolates the system geometry from
  square to triangular type. Four regimes are identified:
  paramagnetic metal, N\'eel-AFM insulator, 120$^{\circ}$-AFM
  insulator, and nonmagnetic insulator (NMI). 
  The phase diagram is based on calculations on a
  $L_{c}=6$ cluster without hopping variation.}
\label{PhaseDiagram-ZeroT}
\end{figure}

Figure~\ref{PhaseDiagram-ZeroT} summarizes the zero-temperature phase
diagram as a function of anisotropy and interaction strength as
obtained by our VCA calculations. 
As stated before, the geometrical frustration is parametrized
by the size ratio $t^{\prime}/t$ of the horizontal and diagonal hoppings in
Fig.~\ref{fig:anisotropic-lattice}.  
In the strong correlation limit, Eq.~(\ref{eq:ham_anisotrop}) is
equivalent to the Heisenberg model, and the system develops the
N\'eel-AFM state at $t^{\prime}/t=0$~\cite{RevModPhys.63.1} and the
120$^{\circ}$-AFM state at $t^{\prime}/t=1$~\cite{PhysRevB.50.10048,
  PhysRevLett.82.3899, PhysRevLett.99.127004}. 
Varying the anisotropy strength, {\it i.e.}, changing the value of
$t^{\prime}/t$, we observe a transition between these two
magnetically ordered states.
The N\'eel-AFM state is surprisingly stable against geometric
frustration, which, among other methods, has also been found in more
sophisticated functional 
renormalization group studies~\cite{ronnyjohannes}: 
Within the VCA, $t^{\prime}/t$ has to be larger than 0.89 in order to destroy the
collinear antiferromagnetic order to establish the 120$^{\circ}$-AFM state.
In contrast, in the weak-coupling limit, the geometric frustration
plays a much more significant role already for small $t'/t$, as it stabilizes the metallic phase in the entire range of
$t^{\prime}/t>0$ for sufficiently small $U/t$.  
(On the square lattice and with nearest-neighbor hopping, due to the Fermi surface nesting,
the ground state of the system is N\'eel-ordered already for
infinitesimal Hubbard $U$. The small offset found in
Fig.~\ref{PhaseDiagram-ZeroT} is a minor finite-size artefact in the VCA.)  
With the increase of $t^{\prime}/t$, the metallic state is stabilized
and extends up to larger values of $U/t$.
For small $t^{\prime}/t$, the metal-insulator transition (MIT) 
coincides with the development of antiferromagnetic order.  

With larger anisotropy, the MIT as a function of coupling strength
acquires a different character. 
In Fig.~\ref{PhaseDiagram-ZeroT}, we observe a transition from a
metallic ``phase'' to a nonmagnetic insulating (NMI) phase.
For approximately $5.2<U/t<9$, the system opens up a charge gap
without developing long-range magnetic order when $t^{\prime}/t>0.7$
(see Fig.~\ref{PhaseDiagram-ZeroT} for the precise boundaries of the
metal-NMI and NMI-magnetic transition). 
The NMI phase is the natural regime where one or several kinds of
quantum spin-liquid (QSL) phases might be located. 
Strong geometric frustration combined with charge fluctuations suppresses the magnetic
ordering in this coupling region. Note that since our numerical methods are
adjusted to the computation of single-particle quantities, it is
impossible for further analyze the specific properties of the NMI
phase, which is indispensable to making concrete predictions for the
expected spin-liquid states.

The appearance of the NMI phase in the intermediate-coupling region
qualitatively agrees with other theoretical
investigations~\cite{PhysRevLett.97.257004,PhysRevLett.97.046402, PhysRevLett.105.267204,
  PhysRevB.75.033102, Sahebsara2008,
  Watanabe2006,Watanabe2008,PhysRevB.87.035143,Dayal2012,Yoshioka2009}. 
In these works and the present study,  the emergence of the NMI phase is
consistently shown to be due to the competition between electronic
correlation and geometric frustration. However, the size of the NMI
phase differs. For VCA, this is partially due to the fact that the lower bound of the
NMI phase slightly depends on the cluster size.
The metal to NMI phase transition occurs at $U/t=5.4$ for
$L_c=3$, $U/t=5.25$ for $L_c=6$, and $U/t=6.3$ for $L_c=12$.
As electrons gain more mobility in a larger cluster, the kinetic
energy of the ground state will become lower in this case. 

Recently, we realized that for hexagonal lattices in particular, this mobility enhancement can be efficiently simulated by introducing 
another variation parameter, {\it i.e.}, $t$, to the VCA
procedure~\cite{Laubach2014a}, (see also Sec.~\ref{variation_t}). 
Hopping $t$ describes the itinerancy of a single electron.
The variation of $t$  thus allows us to minimize the kinetic energy, which
largely recovers the same physics in a small cluster that would
emerge in a larger cluster.
In contrast, the upper bound of the NMI phase, {\it i.e.} the NMI to
120$^{\circ}$-AFM phase transition boundary, is determined by the
collective behavior of all electrons in the system, and thus is less
affected by the variation of single-electron hopping. 
As a result, we find that, by varying $t$, the lower bound of the NMI phase
becomes $U/t=7.5$ for the isotropic triangular lattice with $L_c=6$,
while the upper bound $U/t=9.4$ is unchanged. This is affirmed by
calculations on a larger cluster $L_c=12$, where we find the MIT at
$U/t=7.2$. 

In our VCA calculations, the MIT boundary is
determined by the opening of the single-particle gap $\Delta_{sp}$.
It is directly obtained from the poles of the Green's function with
nonzero weight. As such, no broadening of the spectral function and
further extrapolation are employed; see
Sec.~\ref{spectra_nobroadening} for more technical details. 
This allows us to accurately determine the charge gap size from the
energy difference between the top of the valence band and the bottom of the
conduction band. 
\begin{figure}[htbp]
\centering
\includegraphics[width=\linewidth]{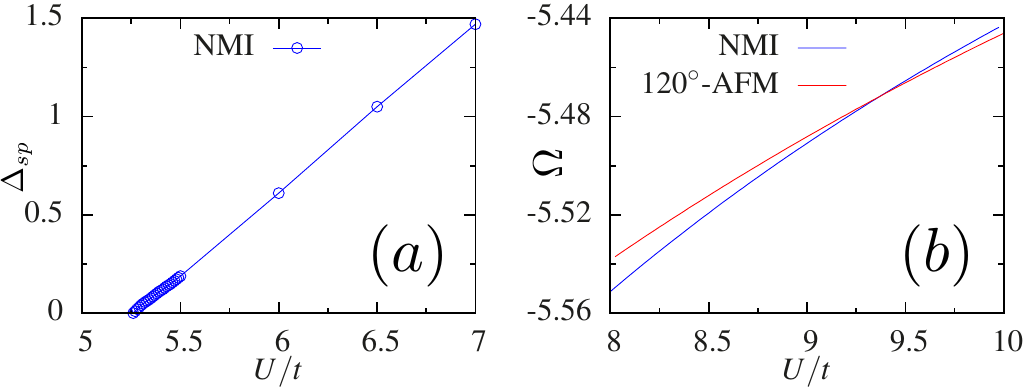}
\caption{ (a) Single-particle gap as a function of the interaction
  strength. 
  The opening of the gap marks the phase boundary of the metal and is
  the lower bound of the NMI phase.  
  (b) The grand potential of the $120^{\circ}$-AFM and the NMI
phases as a function of different interactions for
$t^{\prime}/t=1.0$. The crossing of the grand potential of different phases marks
the transition and indicates its first-order nature.}
\label{Grand-Potential}
\end{figure}
We find, as displayed in Fig.~\ref{Grand-Potential}(a), that the
charge gap $\Delta_{sp}$ opens at $U=5.25$ for $t^{\prime}/t=1$,
indicating a MIT
at $U_c=5.25$. In addition, we also determined the boundary between
different 
magnetic phases as well as the nature of the phase transitions from
the comparison of the VCA grand potential for different phases in
Fig.~\ref{Grand-Potential}(b). 
Around a transition, the preferred phase possesses the lower grand-potential energy, and the transition
is characterized by the crossing of the grand-potential energy of different phases.
In Fig.~\ref{Grand-Potential}(b), we show an example of the 
$120^{\circ}$-AFM to NMI transition for $t^{\prime}/t=1.0$.
The two VCA grand-potentials cross at $U/t=9.4$, which indicates that
this transition is of first order. 
(If the two grand potentials smoothly change
from one to another without any crossing, the transition is of second order or higher.) 

\begin{figure}[htbp]
\centering
\includegraphics[width=\linewidth]{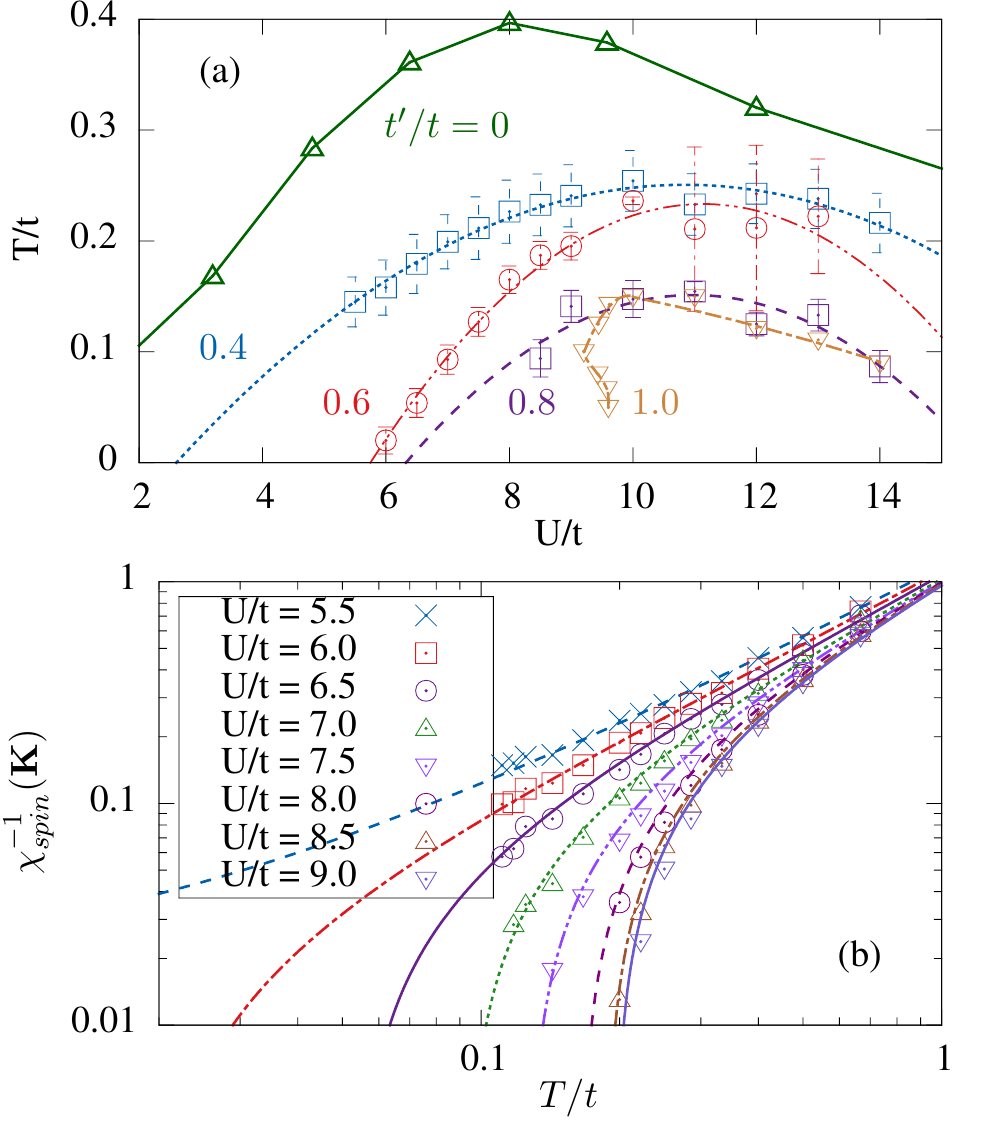}
\caption{(a) N\'eel temperatures
  as a function of interaction $U/t$ for five different anisotropy
  strengths $(t^{\prime}/t)$. The maximum N\'eel temperature appears to
  shift to higher $U/t$ values as a function of anisotropy.
  The error bars represent the uncertainty of the numerical
  extrapolation. 
  The $t^{\prime}/t=1.0$ curve is a replot of the result from
  Ref.~\onlinecite{PhysRevB.83.085102}.
  (b) Example for the determination of the N\'eel temperature via the
  extrapolation of the inverse spin susceptibility for
  $t^{\prime}/t=0.6$. 
}
\label{Neel}
\end{figure}

\subsection{LDFA}

The stabilization of the metallic state due to geometric frustration
can also be seen from the finite-temperature LDFA calculations displayed 
in Fig.~\ref{Neel}(a).
There, the N\'eel temperatures are displayed as a function of
interaction for different anisotropy strengths. 
Strictly speaking, there is no finite-temperature magnetic transition
in two dimensions, according to the Mermin-Wagner
theorem~\cite{PhysRevLett.17.1133}.  
The transition still appears in a method that includes certain implicit
IR cutoffs, 
such as that given by the partial mean-field character in DMFT and LDFA. 
The numerical finding is useful anyway, as
the magnetic correlations are correctly described in this type of
calculations.  
Furthermore, the finite-temperature magnetically ordered phase can be
realized in a slab of multilayer triangular systems, where the mean
field character of DMFT and LDFA can mimic the generic effect of three-dimentional coupling. 
In our study, the
N\'eel temperatures are obtained from the extrapolation of the 
inverse spin susceptibility.
Following Eq.~(\ref{Eq:exact-2}), we calculate
$\chi_{Q}^{s}(i\Omega_{m}=0)$ in the entire first Brillouin zone (BZ)
and extrapolate the inverse of the leading value of the
temperature-dependent $\chi_{Q}(i\Omega_{m}=0)$ to $T/t=0$.  
The N\'eel temperature is given by the temperature where the
extrapolated $\chi_{Q}^{-1}(i\Omega_{m}=0)$ becomes zero.
Examples of the spin susceptibilities and the extrapolation can be
found, in Fig.~\ref{Neel}(b) for $t^{\prime}/t=0.6$. 

As shown in Fig.~\ref{Neel}(a), increasing the anisotropy strength
greatly suppresses the N\'eel temperature, especially in the
weak-coupling region.
At $t^{\prime}/t=0$, the square lattice is recovered and the N\'eel
temperature is non-zero for any finite interaction ($U_c=0$), indicating that
long-range magnetic correlations are well established for $U/t > 0$. 
Increasing $t^{\prime}/t$ results in the suppression of the N\'eel
temperature; as a result $U_{c}$ increases.
This coincides with what is obtained from the $T=0$ VCA calculations shown in
Fig.~\ref{PhaseDiagram-ZeroT}. 
Still, we cannot expect an exact quantitative agreement on $U_{c}$ from these two methods, because both of them are subject to certain approximations.
On the current level of approximation (see Sec.~\ref{Sec:Method} for more
details about the reference system size and the self-energy diagrams
considered), Figs.~\ref{Neel}(a) and ~\ref{PhaseDiagram-ZeroT}
converge to the same conclusion, that the enhanced spatial anisotropy
stabilizes an extended metallic phase and suppresses the propensity to
magnetic ordering.

\begin{figure}[btbp]
\centering
\includegraphics[width=\linewidth]{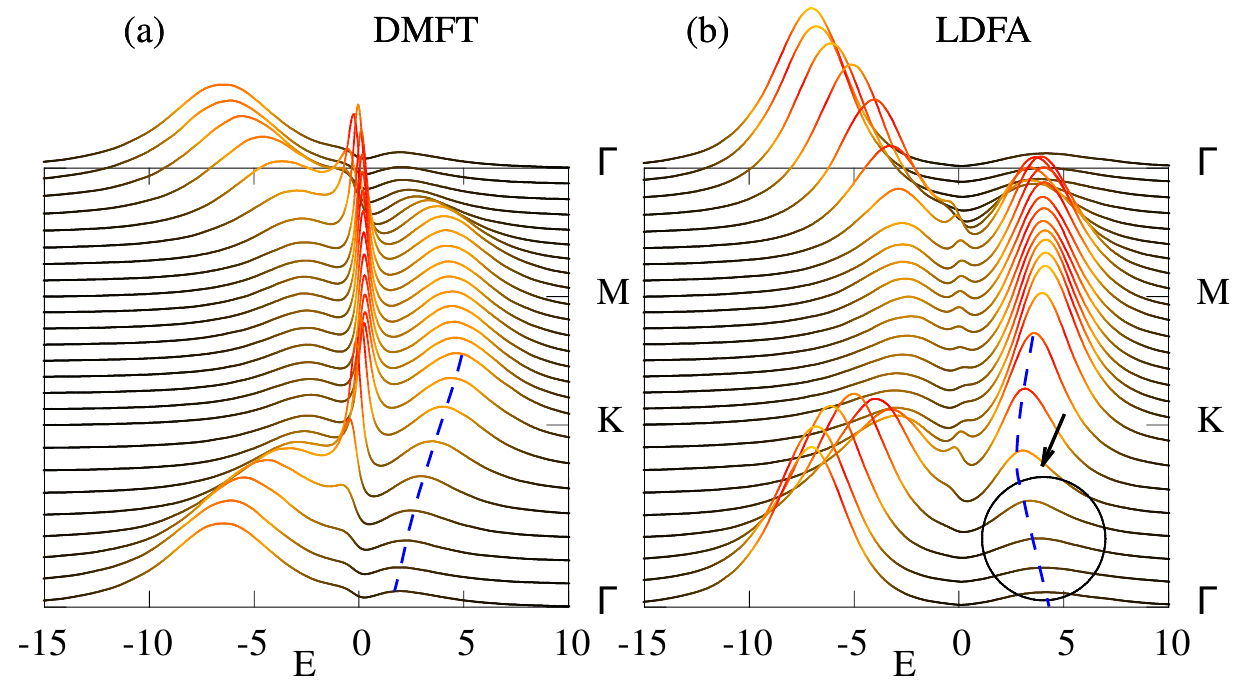}
\caption{Momentum dependence of the spectra for the triangular Hubbard
model calculated from (a) DMFT and (b) LDFA with anisotropy strength
$t^{\prime}/t=0.8$ at $T/t = 0.154$ and $U/t = 9.0$. Due to the
development of  antiferromagnetic correlations, a shadow
band appears above the Fermi level around the $\Gamma$ point, as shown in
the circle in (b). The arrow indicates the symmetry shift induced by the appearance of the shadow band.
This shadow band is absent in the DMFT calculation. The coherent
peak at the Fermi energy $E=0$ is not suppressed due to the missing of
non-local correlations.}
\label{Akw}
\end{figure}

When the system approaches the antiferromagnetic state as a function
of $U$ or $T$, the magnetic correlations drive the spin
susceptibility divergently but also leave fingerprints in the single-particle spectra. 
In the case of the square lattice, with the development of
commensurate $Q=(\pi, \pi)$ antiferromagnetism, the effective magnetic unit cell becomes
twice the size of the original unit cell.  
The single-particle spectra then pick up the new symmetry associated
with the magnetic unit cell, which results in a ``shadow band'' around the
$\Gamma$ point~\cite{PhysRevB.39.11663}. 
In the fully isotropic triangular case, the magnetic correlation is of the
$120^{\circ}$ type. The resulting magnetic unit cell is then three
times the size of the original unit cell. 
The original band will further be folded with respect to the magnetic
zone boundary, which will also generate a shadow band around 
$\Gamma$. 
Thus, detecting the appearance of the shadow band can help us
track the magnetic correlations of the system from the analysis of
single-particle spectra.

Comparing DMFT and LDFA data is
instructive to highlight the additional non-local corrections kept in
LDFA. A detrimental problem of the DMFT lies in the local approximation,
making it
incapable of describing long-range correlations. Thus, the shadow band
induced by the magnetic correlation should be less obvious or even
absent in a DMFT calculation. 
In Fig.~\ref{Akw}, we show the single-particle spectra for
$t^{\prime}/t=0.8$, $U/t=9$, and $T/t=0.154$. 
The LDFA calculations were performed on the Matsubara axis, and the
transformation to the real-frequency axis is accomplished by using the
stochastic analytical continuation~\cite{Beach}. 
The chosen temperature is slightly above the N\'eel temperature, at which
the magnetic correlations begin to fully unfold. 
We thus would expect new folded bands to appear in the single-particle
spectra. 
In Fig.~\ref{Akw}(b), as indicated in the circle, there is a band with
less intensity that develops around $\Gamma$. 
This band is absent in the tight-binding model of
Eq.~(\ref{eq:ham_anisotrop}) and is a direct result of the band
backfolding with respect to the magnetic zone boundary.
The symmetry shift induced by the enlarged magnetic unit cell is
indicated by the arrow in Fig.~\ref{Akw}(b).
This  shadow band locates at a finite energy at the $\Gamma$
point, resulting in a slope change of the band above the Fermi level, as
indicated by the blue dashed line in Fig.~\ref{Akw}(b). 
This magnetic-ordering-induced shadow band is not correctly resolved
in the DMFT calculation. 
As displayed in Fig.~\ref{Akw}(a), the band close to $\Gamma$ is not a
band folded from the emergent symmetry, {\it i.e.}, the $120^{\circ}$
symmetry; it is only the reminiscence of the band between 
$M$ and $K$. It gradually approaches zero energy at the
$\Gamma$ point, as can be seen by following the blue dashed line. 
Another clear difference between the DMFT and the LDFA results lies in
the suppression of the quasi-particle peak at the Fermi level. 
This is again due to the non-locality missing in the local
approximation of the DMFT, which is accurately kept in LDFA.

\begin{figure}[ttbp]
\centering
\includegraphics[width=\linewidth]{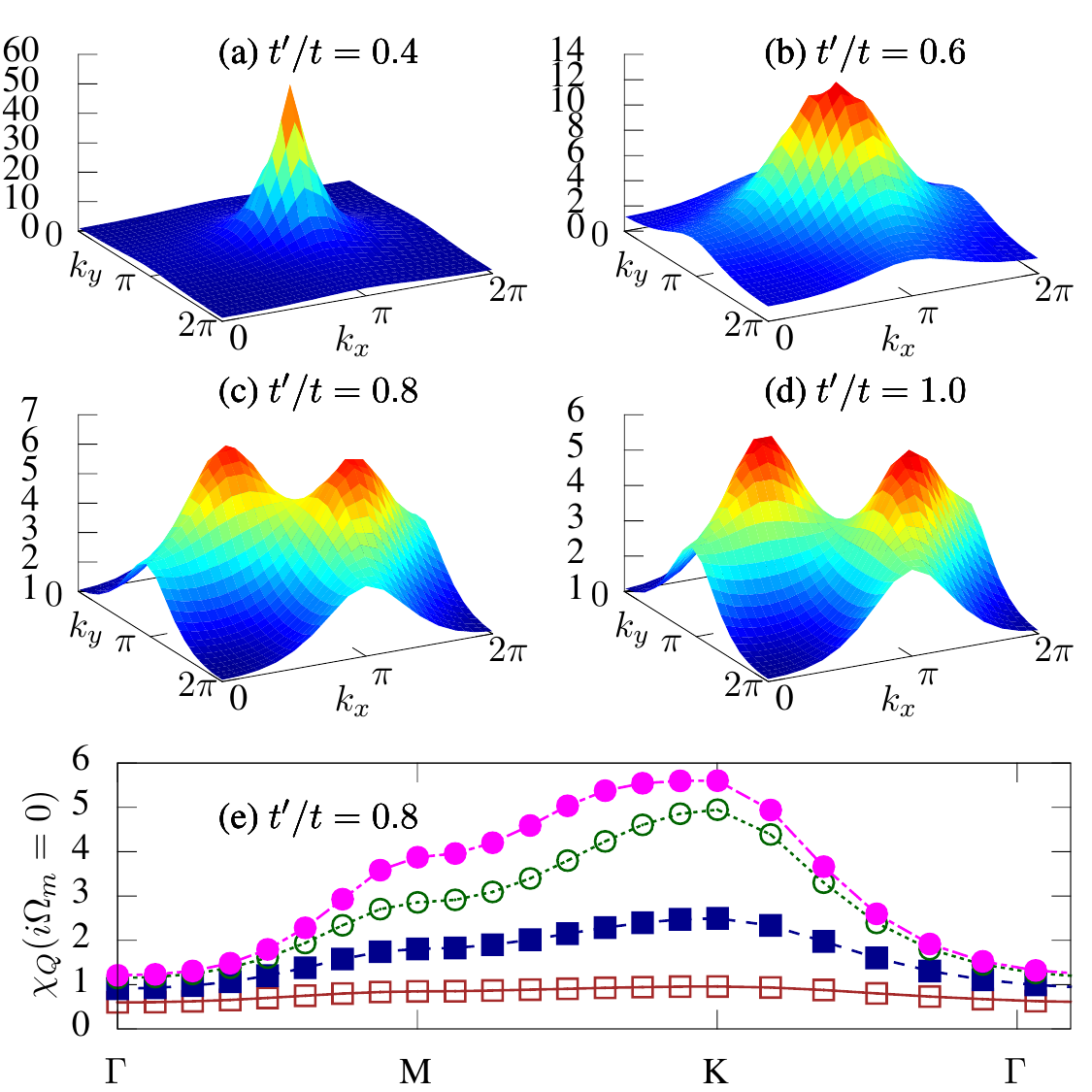}
\caption{(a)-(d) Spin susceptibilities $\chi_{Q}(i\Omega_{m}=0)$ for
  $T/t \sim 0.286$ and $U/t = 9$ with different anisotropy
  strengths. 
  The temperature is chosen to be low enough that all spin
  susceptibilities remain finite.
  The divergence of $\chi_{Q}(i\Omega_{m}=0)$ indicates the
  development of magnetic order with the corresponding $\vec{Q}$.
  (e) Formation of the antiferromagnetic
  correlation with the decrease of temperature for $t^{\prime}/t=0.8$
  and $U/t=9$. 
  From bottom to top, temperatures vary as $T=1.0, 0.5, 0.33, 0.25$.} 
\label{Sus}
\end{figure}

The transition between the two types of magnetic order in
Fig.~\ref{PhaseDiagram-ZeroT} is also observed at finite temperature.
Figure~\ref{Sus} shows the momentum-dependent susceptibilities
$\chi_{Q}(i\Omega_{m}=0)$ in the entire first BZ at $U/t=9$ and $T/t\sim 0.286$, 
for four representative values of $t^{\prime}/t$. 
As discussed in Sec.~\ref{Sec:Method}, the instability of the
paramagnetic solution reveals the formation of a magnetically ordered phase. 
As shown in Fig.~\ref{Sus}(a) and Fig.~\ref{Sus}(b), with smaller anisotropic
strengths, $\chi_{Q}(i\Omega_{m}=0)$ shows a single-peak located at the N\'eel-AFM
$Q=(\pi, \pi)$.
The increase of $t^{\prime}/t$ results in peak broadening as for
the case of $t^{\prime}/t=0.6$ depicted in Fig.~\ref{Sus}b. 
A further increase of the anisotropy strength splits the peak, and
generates a double-peak structure of the spin susceptibility, which
highlights the evolution from N\'eel-AFM to the $120^{\circ}$-AFM. 
At $t^{\prime}/t=1.0$, the valley between the two peaks at $(2\pi/3,
2\pi/3)$ and $(4\pi/3, 4\pi/3)$ becomes even deeper.
It agrees with our VCA results (Fig.~\ref{PhaseDiagram-ZeroT}). 

At finite temperatures, the LDFA allows us to analyze the
thermodynamics of the Hubbard model on the anisotropic triangular lattice. In one of
our recent works~\cite{PhysRevB.89.161118}, 
by using LDFA, we showed that thermodynamical quantities such as the
entropy substantially enhance our understanding of the competing roles of
geometric frustration and electronic correlations.
We found that geometric frustration favors the effect of ``adiabatic
cooling"; that is, following a constant entropy curve, increasing the
interaction results in an effective decrease in temperature. 
This is in contrast to the situation in
square~\cite{PhysRevLett.104.066406} and
cubic~\cite{PhysRevLett.107.086401} lattices, where the entropy is nearly
a constant function of interaction in the weak-coupling region.  
The effect of ``adiabatic cooling" has been found for the
honeycomb-type lattice~\cite{PhysRevLett.109.205301}. We speculate
that the
geometric frustration as imposed by the triangular lattice enhances the decrease
in entropy as a function of interaction strength, which is discussed
in the following.
\begin{figure}[btbp]
\centering
\includegraphics[width=\linewidth]{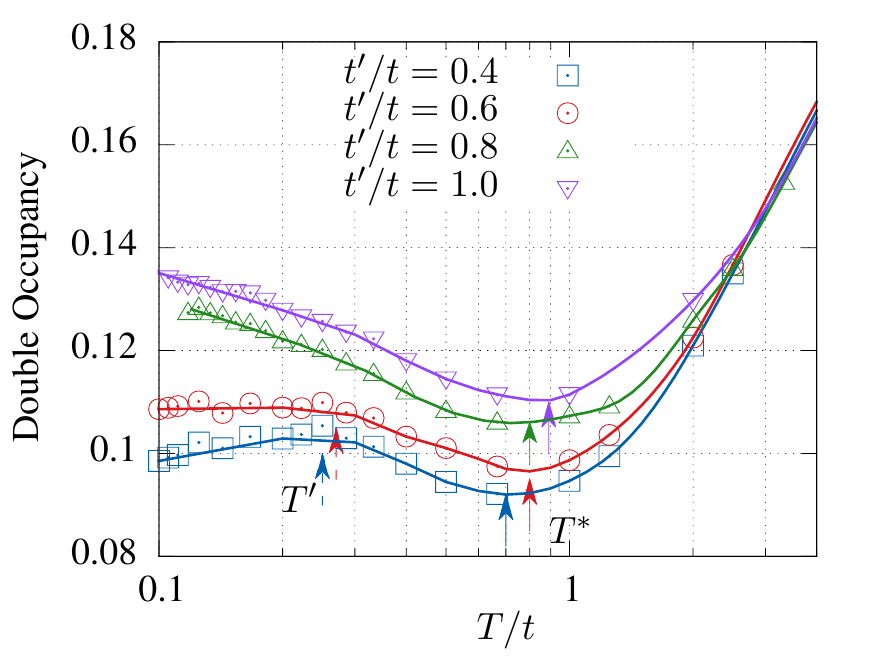}
\caption{The double occupancy $D$ displays an individual temperature dependence
  for different anisotropy strengths. $U/t=6$, and 
for all temperatures studied in LDFA, the system remains metallic for
this interaction strength.}
\label{DouOcc}
\end{figure}

Figure~\ref{DouOcc} displays the double occupancy $D=\langle
n_{\uparrow}n_{\downarrow}\rangle$, whose low-temperature
behavior reveals more information than just the degree of electron localization. 
The double occupancy relates to the entropy $S$ through a Maxwell
relation via the Hellmann-Feynman theorem, {\it i.e.} 
\begin{equation}\label{DouOcc-S}
(\frac{\partial S}{\partial U})_{T, V}=-(\frac{\partial D}{\partial
    T})_{U, V}. 
\end{equation}
As discussed in Ref.~\onlinecite{PhysRevB.89.161118}, the negative-entropy
slope for temperatures lower than a characteristic temperature $T^{*}$
indicates that the entropy will increase with the increase in
interactions. 
As shown in Fig.~\ref{DouOcc}, for temperatures smaller than
$T^{*}/t\sim 0.8$, in the isotropic triangular case ({\it i.e.},
$t^{\prime}/t=1$), $D$ decreases upon increasing $T$~\cite{PhysRevB.89.161118}.
It becomes more intuitive by rewriting Eq.~(\ref{DouOcc-S})
as 
\begin{equation}
\frac{C_{v}}{T}\left(\frac{\partial T}{\partial U}\right)_{S} =
\left(\frac{\partial D}{\partial T}\right)_{U},
\end{equation}
where $C_{v}$ is the specific heat. This immediately implies that,
keeping the entropy constant, an increase in $U$ results in a decrease in
$T$ for $T\le T^{*}$.
The influence of frustration effects becomes clear in Fig.~\ref{DouOcc}.
With the reduction in anisotropy, the negative slope of the double
occupancy below $T^{*}$ becomes less obvious.
In the unfrustrated limit, {\it i.e.}, $t^{\prime}/t=0$, the double
occupancy would resemble that in a square lattice, indicating that no
``adiabatic cooling" is possible. 
At a fixed temperature below $T^{*}$, frustration increases the value
of the double occupancy, resulting in the enhancement of its negative
slope. 
We conclude that frustration is the reason for the
``adiabatic cooling" in the anisotropic triangular system. 
In addition to the change of the slope, we find that the characteristic
temperature $T^{*}$ becomes slightly smaller with the decrease of
$t^{\prime}/t$ (see the solid arrows around $T/t\sim 0.8$ in
Fig.~\ref{DouOcc}). 
In the curves for $t^{\prime}/t=0.4, 0.6$, we also observe a second
characteristic temperature $T^{\prime}$ due to the evolution of system
geometry from square to triangular (dashed arrows around
$T/t\sim 0.25$).  

\section{Conclusions}\label{Sec:Conclusions}

We have conducted a detailed single-particle spectral analysis of the
Hubbard model on the anisotropic triangular lattice for zero and finite
temperatures. Focusing on the role of anisotropy and interaction
strength, we have identified the significant features of the phase
diagram displaying, e.g., a magnetic
transition regime between N\'eel-AFM and 120$^{\circ}$-AFM  orders as
well as, in particular, a non-magnetic insulating regime. Once
set at an anisotropy value near the triangular limit, the NMI
domain quickly broadens in terms of the range of
$U/t$ as a function of anisotropy, along with a more extended
metallic regime for weaker coupling. It is exactly the NMI regime which
might prove the most relevant for unconventional organic
compounds, as one or several spin-liquid
phases can potentially appear in this window of parameter space. While
this question is beyond the framework of the current investigation,
which focused on single-particle quantities, it will be
worthwhile to follow up on the identification of the NMI regime and to adapt methods
capable of calculating multiparticle vertices in order to identify
the nature of the quantum many-body phase. Along this path, 
our VCA/LDFA study is helpful in that it constrains the
interesting parameter window to be scanned by other approaches such as
variational Monte Carlo methods.
At finite temperature, in line with our $T=0$ study, we find that the
anisotropy substantially suppresses the magnetic ordering of the system.
The formation of the shadow band at the $\Gamma$ point in the LDFA
calculations shows that, going beyond DMFT, the LDFA is capable of
describing magnetic effects due to the inclusion of non-local
correlations.  
A characteristic temperature $T^{*}$ is identified in the
double occupancy, below which the double occupancy decreases
upon increasing temperature.
This opens up the possibility of ``adiabatically cooling" the system by
increasing the interactions while keeping the entropy constant. 
We find that the geometrical anisotropy favors ``adiabatic cooling", that is,
increasing the anisotropy results in a larger negative slope of the
double occupancy below $T^{*}$.

{\it Note added.} Recently, we
became aware of a related VCA study
of the anisotropic triangular Hubbard model by
Yamada~\cite{PhysRevB.89.195108}. The subset of VCA findings contained in
our paper agrees with this study.    

\section{Acknowledgement}
W.H. and R.T. acknowledge the hospitality of the KITP where some of
this work was finished and supported by the workshop program NSF PHY11-25915. 
G.L., M.L. and W.H. acknowledge the
DPG Grant Unit FOR1162. R.T. and W.H. are supported by the European Research
Commission through ERC-StG-TOPOLECTRICS-336012. We thank LRZ
Munich for generous allocation of CPU time.
  
\appendix
\section{Technical refinements and caveats of the VCA}
\label{AppA}
In this appendix, we present  useful technical
improvements to the VCA method for the current study. 
They concern the efficient simulation of more nonlocal fluctuations in a
small reference cluster, and the precise determination of the
spectral function from the pole structures of the single-particle
Green's function. Furthermore, we illustrate the importance of an
appropriate choice of reference cluster. Specifically, we show that
the analysis of superconductivity is heavily affected by the discrete
rotation symmetries of the reference cluster, rendering the VCA
approach inaccurate for a reliable investigation of superconductivity
for the anisotropic triangular lattice.

\subsection{Variation of single-particle hopping}\label{variation_t}
The variation of single-particle hopping $t$ in VCA is usually not
important for the study of strongly correlated systems, as its
influence is negligibly small. 
We recently found, however, that this effect becomes
important for the Hubbard model on the honeycomb lattice with small
and intermediate correlations~\cite{Laubach2014a}, which is also in
line with previous studies on the square lattice \cite{dahnken-2004-70,Aichhorn2004a}.  
In the current study, this is the regime where the MIT happens, and the
NMI phase emerges in proximity to a metallic domain.  
Thus, we find that the variation of the single-particle hopping $t$ is
crucial for the analysis of the Hubbard model on the anisotropic
triangular lattice. 
 \begin{figure}[bbb]
\centering
\includegraphics[width=0.95\linewidth]{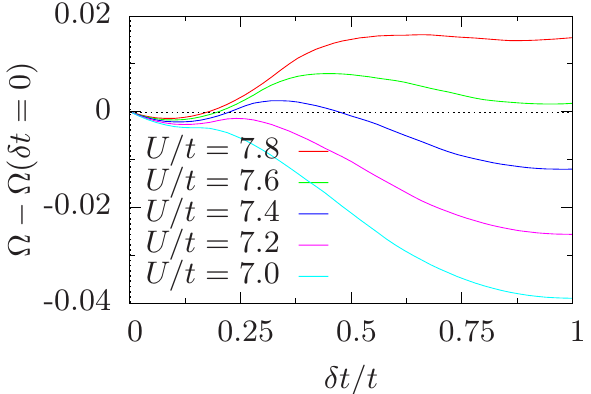}
\caption{VCA grand potential as a function of $\delta t/t$. In the
  metallic phase, {\it i.e.}, $U/t< 7.5$, the minimum is around $\delta
  t/t\sim 1$, representing a strong non-local modification of the
  hopping amplitude inside the reference cluster. 
  This is not seen for the insulating regime where $U/t>7.5$,
  as the minimum of $\delta t/t$ quickly moves to zero. For the two specific
  values of $U/t$ shown in this plot, the minima are at $\delta t/t=0.1$ for $U/t=7.6$
  and at $\delta t/t=0.08$ for $U/t=7.8$. 
  The correction implied by the variation of $t$ becomes negligible in the
  strongly correlated regime.} 
\label{fig:MIT-VCA}
\end{figure}

\begin{figure}[ttt]
  \centering
  \includegraphics[width=0.95\linewidth]{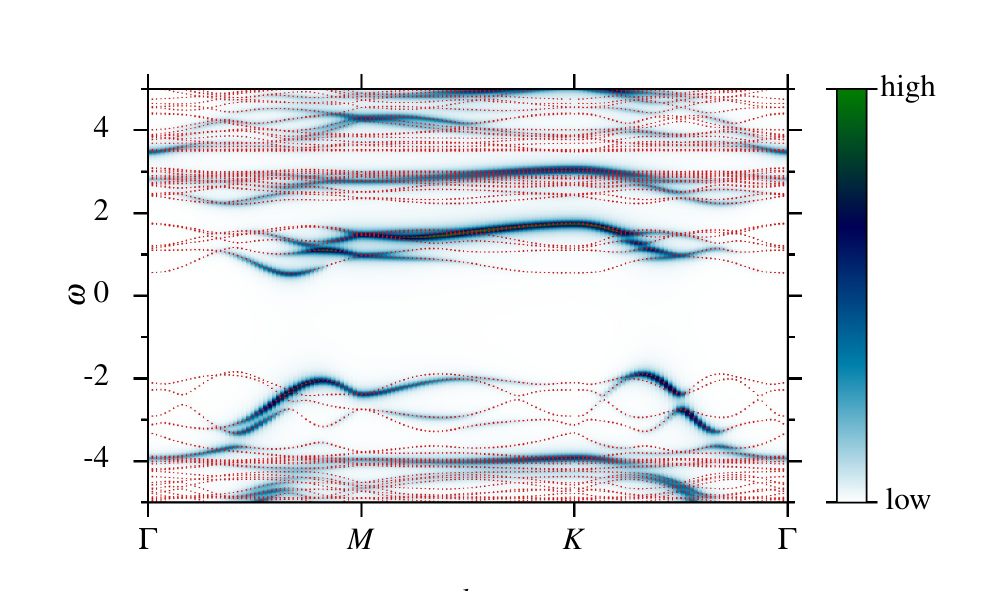}
  \caption{Comparison of the spectral function calculated with a
    phenomenological broadening
    $\eta=0.1$ (density plot with color code) and the exact pole
    structure of the Green's function with non-zero weight (red
    dots).} 
  \label{fig:spec-triang-u8-ex}
\end{figure}

Figure~\ref{fig:MIT-VCA} displays the grand potential as a function of
$\delta t=t'-t$, where $t^{\prime}$ is the optimal value of $t$ that
minimizes the grand potential.  
At $U/t=7.0$, the minimum of the grand potential is at $\delta t/t\sim
1$, representing 
a strong enhancement of the dynamics within the reference cluster, as
$t^{\prime}\sim 2t$. 
The adjusted dynamics leads to an increase in the critical value for 
the MIT to $U/t\sim7.5$. 
The minima of the grand potential move to $\delta t/t=0.10$ and $0.08$
as interactions increase to $U/t=7.6$ and $U/t=7.8$, respectively.  
Figure~\ref{fig:MIT-VCA} clearly shows the trend of $\delta t
\rightarrow 0$ when stronger interactions are present.  
In contrast to the MIT, the magnetic phase transition of the Hubbard
model on the isotropic triangular lattice takes place at even stronger interactions 
$U/t=9.4$, where the variation of the hopping is found to be negligible.  
As a consequence, with the variation of the single-particle hopping
$t$, the NMI phase appears to be in the regime $7.5<U/t<9.4$, which nicely agrees
with other works, some of which employ different approaches~\cite{ PhysRevLett.105.267204,  
  PhysRevB.75.033102,Watanabe2006,Watanabe2008,PhysRevB.87.035143,Dayal2012,Yoshioka2009}.

\subsection{Exact evaluation of the single-particle gap through spectra function without broadening
  factor}\label{spectra_nobroadening}

In VCA, the spectral function as well as the local density of states
(LDOS) is usually calculated from the single-particle Green's  function with an broadening factor $i\eta$. A precise value of the single-particle gap can be only obtained by extrapolating $\eta$ to zero.
 Here, we present a scheme to evaluate the exact single-particle gap
 without any such broadening factor $\eta$. 
  The single-particle Green's  function is calculated \cite{aichhorn:235117} as
\begin{align}
  {G}=&\frac{1}{({QgQ}^\dagger)^{-1}-{V}}
  ={Q}\frac{1}{{g}^{-1}-{Q^\dagger VQ}}{Q}^\dagger \,,
  \label{eq:gittergreens_mitQ}
\end{align}
where ${g}^{-1}=\omega-{\Lambda}$ is a diagonal matrix
and ${\Lambda}_{mn}=\delta_{mn}\omega_m'$ is the excitation
spectrum of the reference cluster. 
The poles of the VCA Green's function ${G}$ are simply the
eigenvalues of the matrix
${M}={\Lambda}+{Q^\dagger VQ}$.
With the diagonal form of ${M}$, {\it i.e.},
${D}_M={S^{-1}MS}$, one can rewrite the VCA Green's function as  
\begin{align}
G=Q\frac{1}{M}{Q}^\dagger=Q\frac{1}{{SD_MS^\dagger}}{Q}^\dagger 
={QS}\frac{1}{{D_M}}{S}^\dagger{Q}^\dagger\,.
\label{eq:gewicht_gittergreen} 
\end{align} 
The weights associated with the poles ${D}_M$ are $({{QS})_{\alpha m}({S^\dagger
    Q^\dagger}})_{m\beta}$. Only poles with nonzero weights 
contribute to the spectral function. 
In Fig.~\ref{fig:spec-triang-u8-ex} we compare the spectra calculated
from Eq.~(\ref{eq:gewicht_gittergreen}) to the ones calculated after
introducing a broadening factor. Clearly, the employment of
Eq.~(\ref{eq:gewicht_gittergreen}) gives rise to much richer
spectra, where some parts in the intensity plot are missing for the
calculations with broadening. 
This new strategy enhances the accuracy of
the VCA method in characterizing the MIT through the single-particle gap as shown in Fig. \ref{Grand-Potential}(a).
 
\subsection{Artificial bias for superconductivity from broken
  symmetries in VCA reference clusters}\label{SC-app}

\begin{figure}[htbp]
  \centering
  \begin{center}
    \includegraphics[width=0.99\linewidth]{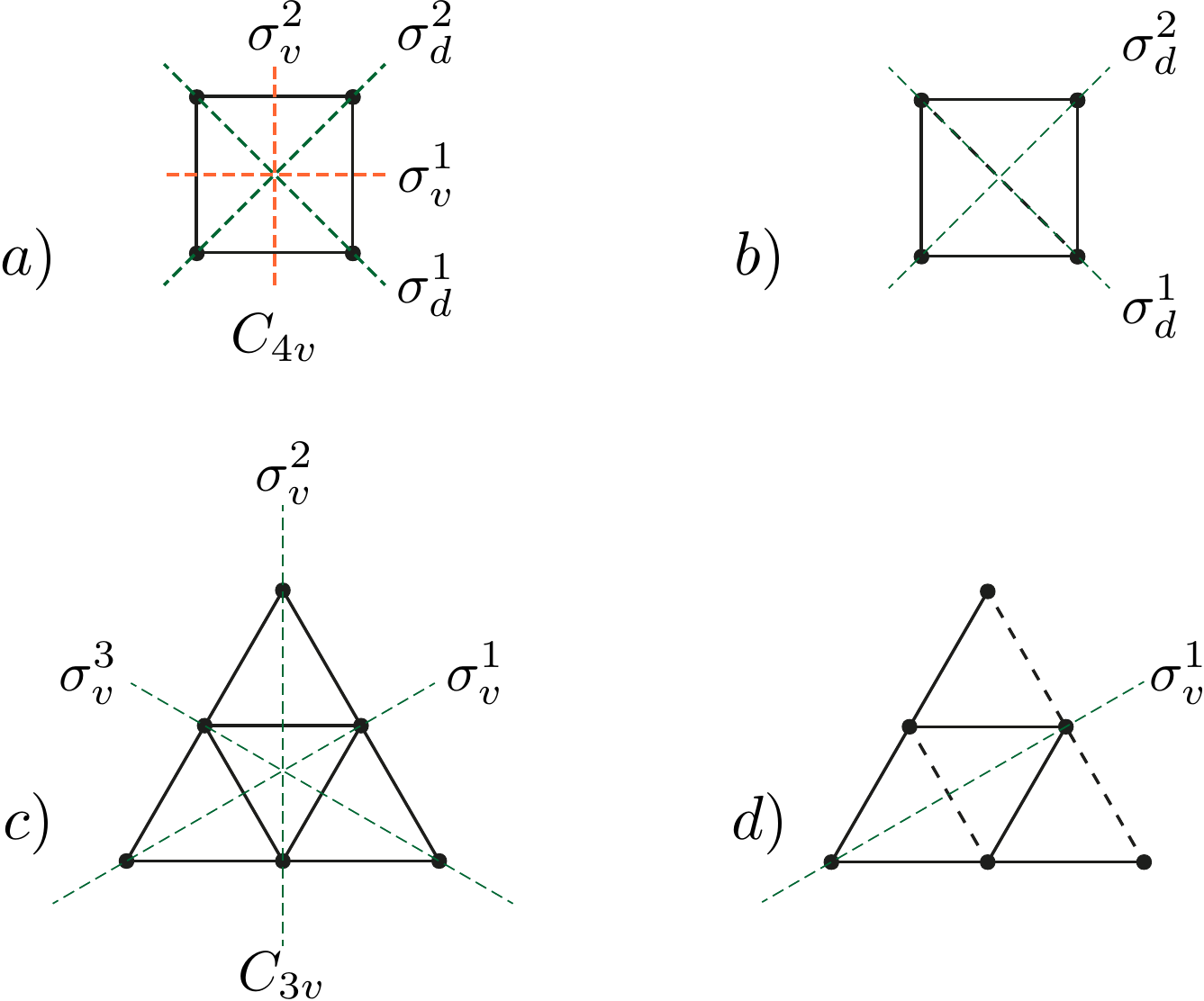}
  \end{center}
  \caption{Cluster symmetries. a) Four-site
    with square-lattice $C_{4v}$ symmetry. b) Anisotropic four-site cluster
    cluster. c) Six-site with triangular-lattice $C_{3v}$ symmetry. (d)
    Anisotropic six-site cluster.}
  \label{fig:cluster-symmetry}
\end{figure}
In the VCA phase diagram Fig.~\ref{PhaseDiagram-ZeroT}, no
superconductivity (SC)  was investigated. In principle, it is
possible in VCA to study SC via appropriate Weiss fields, and indeed, this has been
previously attempted for the Hubbard model on the anisotropic triangular
lattice~\cite{PhysRevLett.97.257004}. In the following, we show that
these previous approaches have to be interpreted with extreme caution, and explicate
why a systematic investigation of SC order for the anisotropic
triangular lattice is not
feasible for VCA or any other finite cluster method as a matter of
principle. Figure~\ref{PhaseDiagram-ZeroT} shall thus be understood as a
tentative phase diagram without the inclusion of SC. For
the isotropic triangular lattice where SC can be
investigated reliably through VCA, we find chiral
$d$-wave SC for a large window from weak to intermediate coupling,
a superconducting solution which was not considered in Ref.~\onlinecite{PhysRevLett.97.257004}.

\begin{figure}[htbp]
  \begin{center}
    \includegraphics[width=0.99\linewidth]{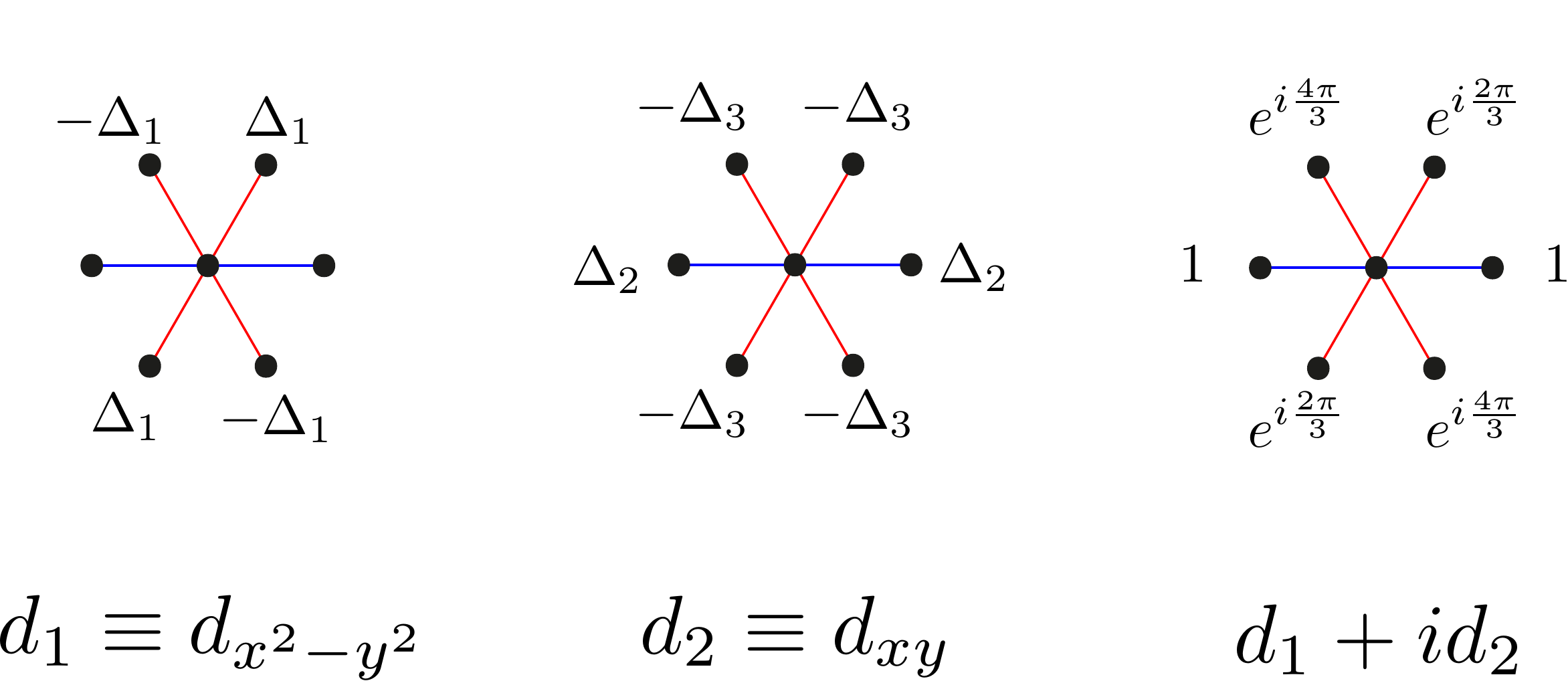}
  \end{center}
  \caption{Real-space superconducting form factors for the anisotropic
    triangular lattice. The amplitude parameters $\delta_i$, $i=1,2,3$,
    are varied in the VCA grand potential. $d_1$ and $d_2$ parametrize
    the SC orders, which become (left) $d_{x^2-y^2}$ and (middle) $d_{xy}$ SC order in
    the isotropic square and triangular limits, respectively. (right) For the
    isotropic triangular lattice, in units of $\Delta_2$, we find
    chiral $d$-wave SC order $d_1+id_2=d_{x^2-y^2}+i d_{xy}$ for
    $\Delta_1=\sqrt{3}/2$ and $\Delta_3=1/2$.}
  \label{fig:formfactors}
\end{figure} 
Figure~\ref{fig:cluster-symmetry} shows the symmetry classification for
the small-size reference clusters we encounter for the anisotropic
triangular lattice. As we intend to interpolate between the square
lattice ($t'/t=0$) and the triangular lattice ($t'/t=1$), we choose a
six-site cluster which is still conveniently tractable numerically and
exhibits commensurability with $C_{4v}$ and $C_{3v}$ in the respective
limits. (As further elaborated on in the main text, this also applies to the
12-site cluster which we, due to the significantly greater numerical
effort used only for special points in the phase diagram.)

In close analogy to magnetic order, SC Weiss fields can be similarly
employed in VCA. The relevant SC form factors are the in-plane $d$-wave
orders $d_{x^2-y^2}$ (Fig.~\ref{fig:formfactors}, left) and $d_{xy}$
(Fig.~\ref{fig:formfactors}, middle), which in total yield three variational SC
amplitude parameters $\Delta_i$, $i=1,2,3$. For $C_{4v}$ symmetry ($t'/t=0$),
both $d$-wave orders are associated with independent one-dimensional irreducible
lattice representations. For $C_{3v}$ symmetry ($t'/t=1$), they form a
single two-dimensional irreducible lattice representation. For generic
$t'/t$, the SC orders are denoted by $d_1$ and $d_2$, respectively.

\begin{figure}[htbp]
  \centering
  \includegraphics[width=0.9\linewidth]{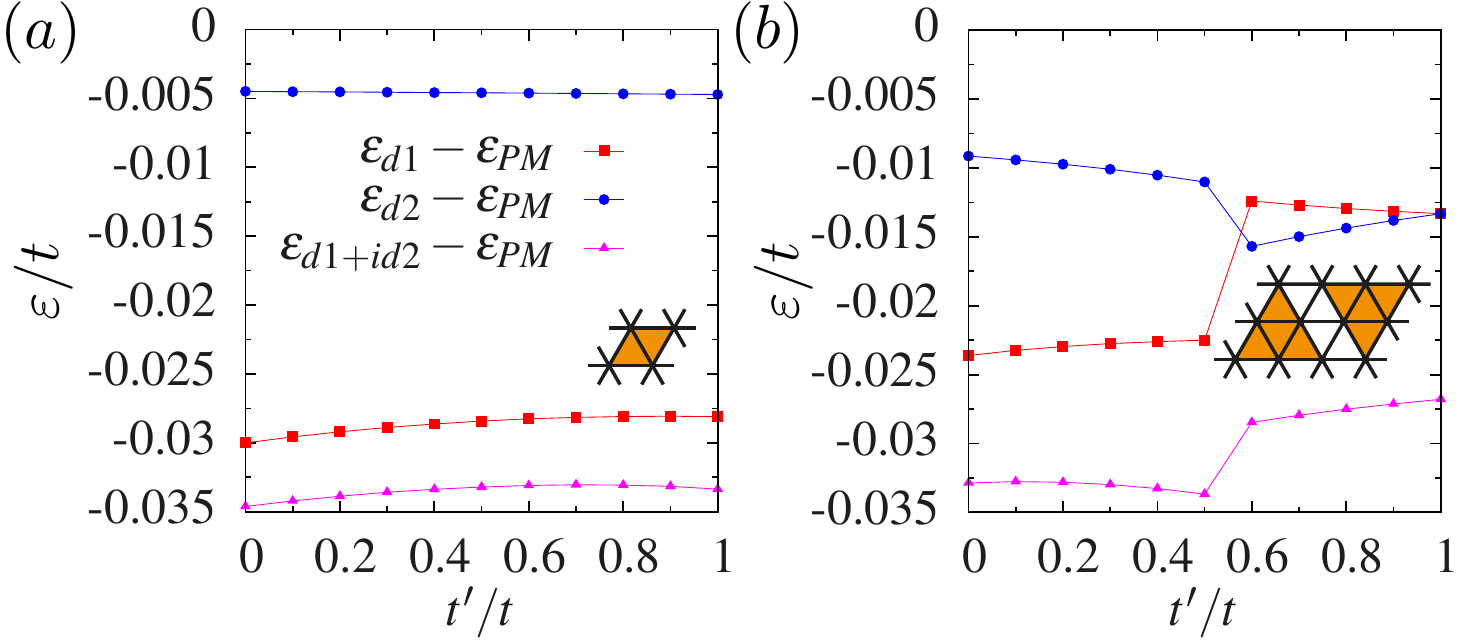}
  \caption{Condensation energy of the (a) four-site and (b) six-site cluster ground
    state for different SC order parameters relative to the
    paramagnetic ground state $\varepsilon_{d_i}-\varepsilon_{PM}$.  
    All energies are taken at $U/t=5$ with fixed respectively chosen SC
    Weiss field $h=0.1t$. $d_1+id_2$ maximizes the
    condensation energy for finite clusters as it removes all
    low-energy spectral weight.}
  \label{fig:condens-energies}
\end{figure}
Let us first analyze the finite-cluster spectra in the presence of the
SC Weiss fields. In Fig.~\ref{fig:condens-energies}, the energy
differences between the paramagnetic ground state and the SC ground
state are plotted as a
function of anisotropy for the four-site and six-site clusters and $U/t=5$. The Weiss
field scale $h/t$=0.1 is big enough that the complex chiral $d$-wave order
parameter $d_1+id_2$ (Fig.~\ref{fig:formfactors}, right) is energetically preferred. This relates
to the fact that only $d_1+id_2$ is fully gapped, while both
individual $d_1$ and $d_2$ retain nodal behavior. (Note, however,
this does not mean that chiral superconductivity should always be
preferred for the infinite VCA system where $h\rightarrow 0$ or for a
finite-cluster spectrum with smaller $h/t$.)

\begin{figure}[htbp]
  \centering
  \includegraphics[width=0.8\linewidth]{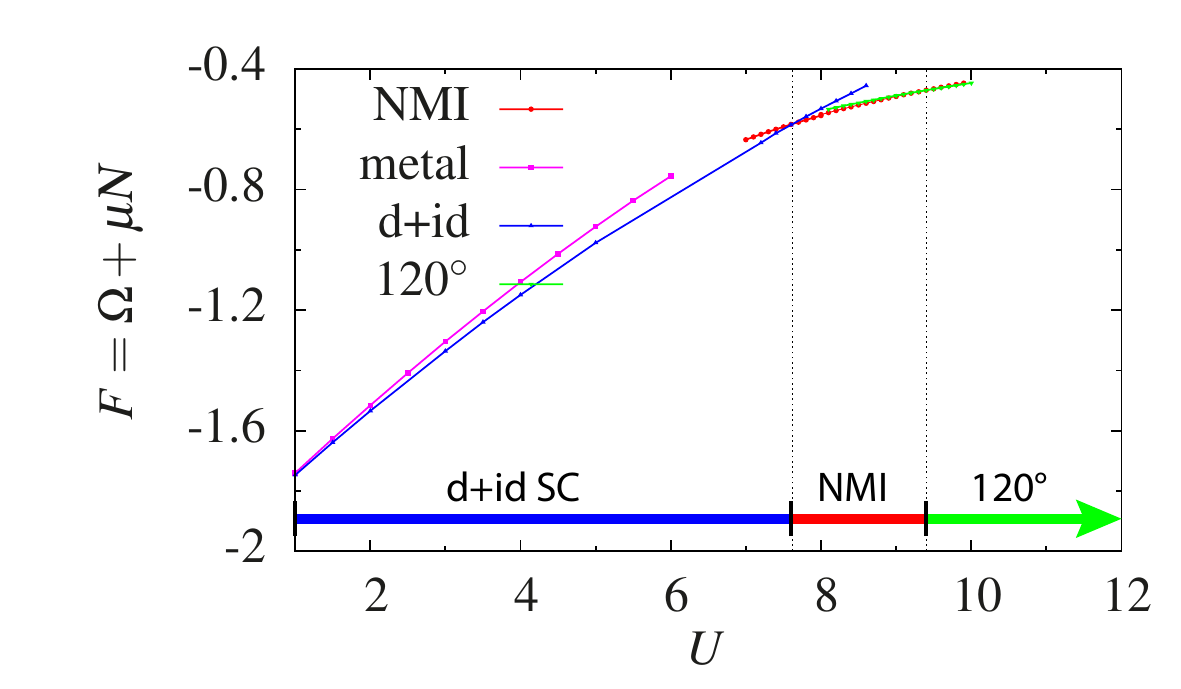}
  \caption{VCA free energy of the Hubbard model on the isotropic
    triangular lattice for a six-site reference cluster. $d+id$ SC
    is stabilized in the weak- to intermediate-coupling regime. It only
    partly overlays the NMI phase obtained in Fig.~\ref{PhaseDiagram-ZeroT}.}
  \label{fig:free-energy-phasediagramm}
\end{figure}
Figure~\ref{fig:condens-energies}(a) shows the four-site cluster spectrum,
which hardly changes as a function of anisotropy. Of the individual
$d$-wave form factors, $d_1 \equiv d_{x^2-y^2}$ has the larger
condensation energy for any anisotropy. This changes as one considers
the six-site cluster [Fig.~\ref{fig:condens-energies}(b)]. Beyond a
certain degree of anisotropy towards the triangular limit, $d_2$ is
preferred over $d_1$. Eventually, $d_1$ and $d_2$ become degenerate for $t'/t=1$, as
dictated by $C_{3v}$ lattice symmetry. Figure~\ref{fig:condens-energies}
demonstrates how the fundamental symmetries of the isotropic triangular
limits are violated by the
four-site cluster, as, e.g., employed in
Ref.~\onlinecite{PhysRevLett.97.257004}. Taking the six-site cluster and
hence accurately accounting for lattice symmetries, the VCA phase diagram
for the isotropic triangular lattice is shown in
Fig.~\ref{fig:free-energy-phasediagramm}. From weak to intermediate
coupling, chiral $d$-wave SC is found, followed by an NMI regime and
$120^{\circ}$-AFM order for increasing $U/t$. The nature of
superconductivity found in our VCA analysis is in accordance with
several approaches such as the functional renormalization group~\cite{PhysRevB.68.104510,PhysRevLett.111.097001,advphys},
parquet renormalization group~\cite{PhysRevB.89.144501}, 
and finite-cluster variational Monte Carlo~\cite{PhysRevB.88.041103}.
Note in Fig.~\ref{fig:free-energy-phasediagramm} that the NMI phase,
as the promising spin-liquid candidate scenario, persists upon the joint
consideration of SC for a sizable coupling regime. 

\begin{figure}[htbp]
  \centering
  \includegraphics[width=0.8\linewidth]{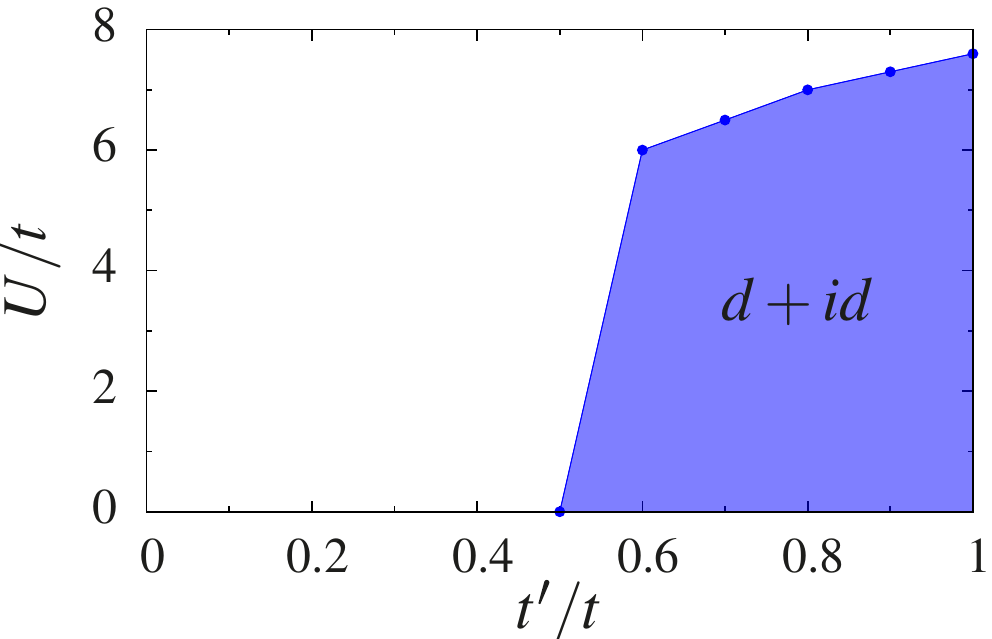}
  \caption{Domain of stable $d+id$ SC grand potential saddle point in
    VCA for the anisotropic six-site reference cluster.} 
  \label{fig:phase-sc-domain}
\end{figure}
How does the SC phase diagram in Fig.~\ref{fig:free-energy-phasediagramm}
evolve for finite anisotropy $t'/t<1$ when $d_1$ and $d_2$ are no longer
degenerate? $d_1+id_2$ dominates as
long as the enhanced gain of the
condensation energy through chiral $d$-wave SC overcomes the energy
splitting between $d_1$ and $d_2$. (Figure~\ref{fig:phase-sc-domain}
shows the domain for which a saddle point of chiral $d$-wave SC is
found in VCA with a six-site reference cluster.) 
Overall, however, the bias from different cluster sizes, as
well as strong finite-size effects in the small clusters in
general, does not allow for a systematic analysis of SC for the
anisotropic lattice. For example, the quantitative analysis of the
transition point between the gaped chiral $d$ wave and nodal $d$ wave does
not appear feasible within VCA: The four-site analysis yields a strong
preference for $d_{x^2-y^2}$-wave SC, while the six-site analysis advocates the chiral
$d$ wave for a large domain of anisotropy. Overall, our findings support the view
that such a question should preferably be addressed through
momentum-resolved approaches in which the adjusted breaking of lattice
symmetries is more accurately accounted for than in a finite-size real-space cluster method.

\bibliographystyle{apsrev4-1}
\bibliography{bibtex}

\end{document}